\documentclass[reprint,prb,showpacs,preprintnumbers,amsmath,amssymb,twocolumns,footinbib,citeautoscript]{revtex4-1}

\usepackage{graphicx}
\usepackage{dcolumn}
\usepackage{bm}
\begin{document}

\title{Resonant transport and electrostatic effects in single molecule electrical junctions}

\author{Carly Brooke,$^1$ Andrea Vezzoli,$^1$ Simon J. Higgins,$^1$ Linda A. Zotti,$^{2,3}$ J. J. Palacios$^{2,4}$  and Richard J. Nichols $^1$}

\affiliation{$^1$Department of Chemistry, University of Liverpool, Crown Street,
 Liverpool L69 7ZD,U.K, $^2$ Departamento de F\'{\i}sica de la Materia Condensada
 and Condensed Matter Physics Center (IFIMAC), Universidad Aut\'onoma de Madrid, 28049
 Cantoblanco, Madrid, Spain, $^3$ Departamento de F\'{\i}sica Aplicada, Universidad de
 Alicante, 03690 Alicante, Spain, $^4$ Instituto Nicolas Cabrera, Universidad
 Aut\'onoma de Madrid, 28049 Cantoblanco, Madrid, Spain}

\date{\today}

\begin{abstract}
In this contribution we demonstrate structural control over a transport resonance
 in HS(CH$_{2}$)$_{n}$[1,4 - C$_{6}$H$_{4}$](CH$_{2}$)$_{n}$SH (n = 1, 3, 4, 6)
 metal \textbar molecule \textbar metal junctions, fabricated and tested
 using the scanning tunnelling
 microscopy-based $I(z)$ method.  The Breit-Wigner resonance
 originates from one of the arene $\pi$-bonding orbitals, which sharpens and moves
 closer to the contact Fermi energy as $n$ increases. Varying the number of methylene
 groups thus leads to a very shallow decay of the conductance with the length of the
 molecule.
 We  demonstrate that the electrical behaviour observed here can be straightforwardly
 rationalized by analyzing the effects caused by the electrostatic balance 
created at the metal-molecule interface.
 Such resonances offer future prospects in molecular electronics in terms of controlling charge transport over longer distances, and also in single molecule conductance switching if the resonances can be externally gated. 
\end{abstract}


\maketitle

\section{Introduction}
The development of reliable techniques for the fabrication and electrical testing
 of metal \textbar single molecule \textbar metal junctions, supported by rapid
 developments in
 theoretical treatment of such junctions, has led to a significant re-invigoration
 of the wider field of ‘molecular electronics’.\cite{nichols2010,chen2007,hybertsen2008,chen2009,cuevas2010} The techniques used can be divided
 into two categories. One category uses solid-state electronics technology to fabricate so-called break junctions, in which a very thin metallic wire is deliberately broken
 (either by electromigration,\cite{park2002} or by the application of mechanical stress
\cite{reichert2002}) in such a way that a molecule can bridge the resulting gap.
 The other category involves use of scanning probe microscopy to fabricate and
 characterise metal \textbar molecule \textbar metal junctions.
 For example, in the scanning
 tunnelling microscopy break junction (STM-BJ) method, a gold STM tip is driven into
 a gold substrate, and then retracted while the current is monitored. \cite{xu2003}
 As the resulting chains of gold atoms are broken, step decreases in the current
 approximately corresponding to the quantum of conductance, G$_{0}$, are seen until the final chain
 of gold atoms is broken. In the presence of molecules that contain terminal potential
 binding groups (e.g. thiols \cite{xu2003}, pyridines
 \cite{xu2003,quek2009,kamenetska2010}, amines\cite{chen2006,venkataraman2006},
 carboxylate anions \cite{chen2006,martin2008}, phosphines\cite{park2007,parameswaran2010}  , thioethers\cite{park2007}), additional much smaller current step decreases
 are seen after the final gold atomic chain breaks, and these correspond to the
 subsequent breaking of metal \textbar molecule \textbar metal junctions.
 Alternatively, in the
 so-called $I(z)$ ($I$, current; $z$, vertical height) method for measuring single
molecule electricla properties, a gold STM tip is brought
 into close proximity of a surface coated with suitable molecules without allowing it
 to come into contact, and is then retracted while the tunnelling current is monitored.
\cite{haiss2003} When a molecule bridges the gap prior to tip retraction, a 
characteristic plateau is seen in the current-distance curve, prior to junction
 breakdown whereupon the current drops sharply. 

\begin{figure}[t]
\begin{center}
\includegraphics[width=8.2cm]{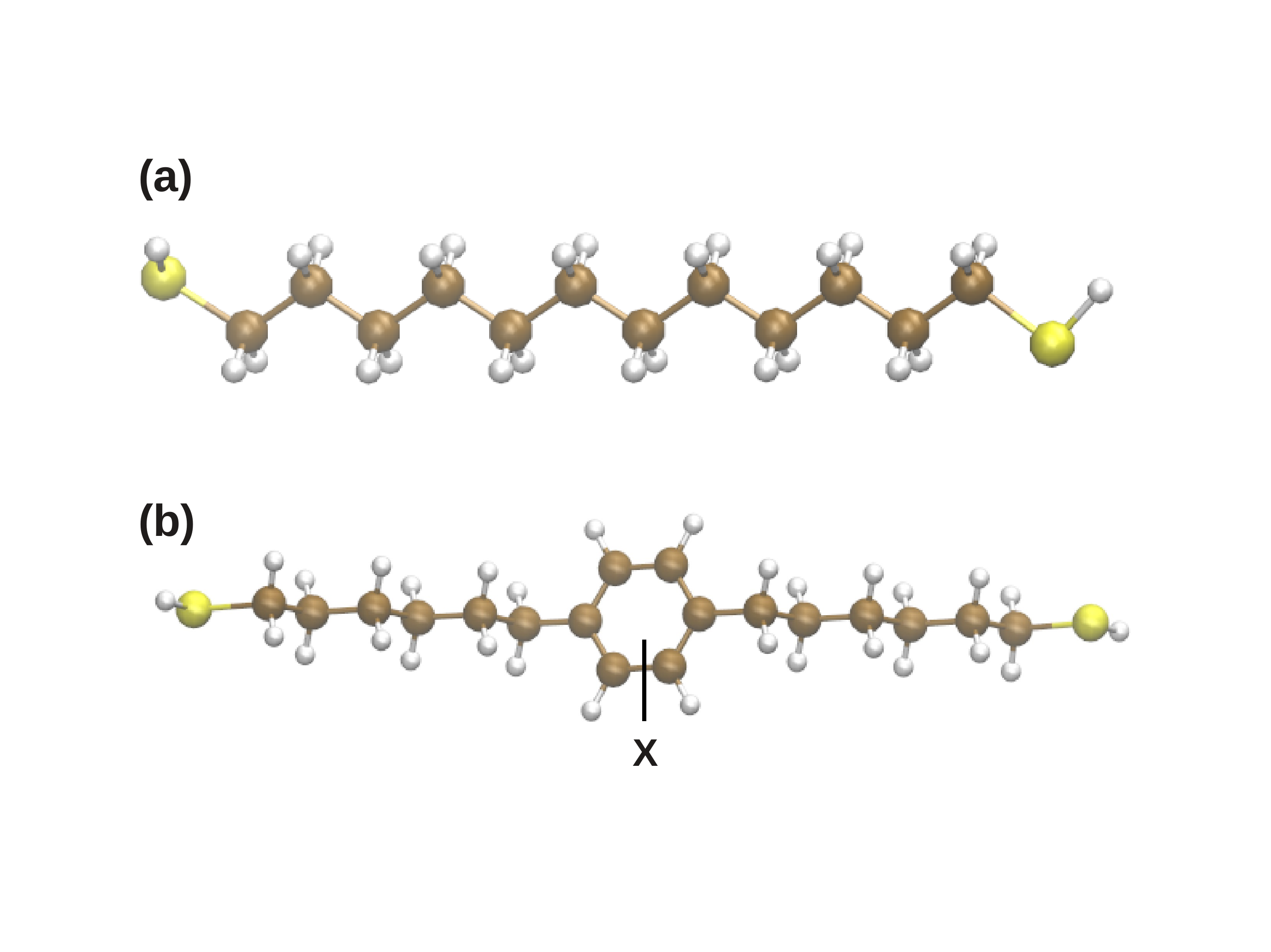}
\caption{Structural comparison of (a) 1,12-dodecanedithiol with (b) 6Ph6 and its
 derivatives (X=2,3,5,6-tetrafluoro-,2,5-dimethyl-2,5-dimethoxy-)}
\label{fig:fig1}
\end{center}
\end{figure}

For the great majority of (relatively short) molecules that have been studied in
 metal \textbar single molecule \textbar metal junctions,
 the mechanism of conductance has been
 found to be coherent tunnelling. \cite{nichols2010} The conductance, G$_{M}$,
 then varies as a function of the length of the bridging molecule according
 to Eq.~(\ref{G-eq1}):
\begin{equation}
G_{M}=G_{con}e^{-\beta R}
\label{G-eq1}
\end{equation}

where G$_{con}=$G$_{0}$T$_{L}$T$_{R}$ (G$_{0}$ is the quantum unit of conductance,
 77.4 $\mu$S, and T$_{L}$ and T$_{R}$ are the transmittances of the left hand and right
 hand connections of the molecule to the metal contacts; G$_{con}$ is called the contact
 conductance and represents the conductance of the system in the absence of the molecular
 backbone), R is the molecular backbone length and $\beta$ is the attenuation factor.
 A plot of lnG$_{M}$ vs. R is used to determine $\beta$.
Often,
 such molecules can be considered as ``passive'' molecular wires, in which the
 conductance decreases with bridge
 length in the simple manner predicted by phase coherent tunnelling. In many cases this
 length decay can be straightforwardly linked to the degree of conjugation, in turn
 related to the proximity of the relevant HOMO or LUMO (``frontier'') orbitals to the
 metal contact Fermi energies E$_{F}$. However, molecular electronics is now aiming to
 achieve more active control over the electrical properties of molecular bridges.
 Approaches to this include electrochemical switching or conformational control of bridge
 structure and electronic states and control of quantum interference in molecular
 circuits.

In earlier work, we showed that the conductance of junctions involving 6Ph6 and its
 derivatives (Fig.~\ref{fig:fig1} ) was much higher (0.7 nS) than that measured for 1,12-dodecanedithiol (i.e. the two alkyl links for 6Ph6 connected back-to-back; 0.028 nS\cite{haiss2009}),
 even though 6Ph6 (2.10 nm S...S distance) is considerably longer than
 1,12-dodecanedithiol (1.72 nm S...S). We suggested that the reason for the unusually
 high conductance of 6Ph6
 was that it behaves as a `molecular double tunnelling barrier' in which a `well' with
 lower HOMO-LUMO ($\pi$-$\pi^{*}$) separation is sandwiched between two `barriers' with
 higher HOMO-LUMO ($\sigma$-$\sigma^{*}$) separation, by analogy with inorganic double
 tunnelling barrier structures involving III-V semiconductors.\cite{Leary2007} Clearly,
 the benzene moiety is capable of acting as an indentation in the tunnelling barrier.
 Interestingly, control over the conductance of the 6Ph6 family could be exerted
 chemically by using substituents on the aryl ring (Fig. ~\ref{fig:fig1});
 electron-donating substituents (Me$-$, MeO$-$) resulted in significantly higher junction
 conductances and the electron-deficient C$_{6}$F$_{4}$ group resulted in a lower
 conductance.\cite{Leary2007}
Many studies of metal \textbar molecule \textbar metal junctions have probed
 the molecular structural
 requirements for obtaining high junction conductance.\cite{ramachandran2003, xu2005,
yamada2008,lee2012, dell2013, sedghi2012, sedghi2011,sedghi2008,li2012,Wang2009}  

The work described in this contribution is aimed at a deeper understanding of
 the evolution of the orbital  alignment in single-molecule junctions.  To this end, we
 have investigated the effect of varying the width of the `barriers' (i.e. the number of
 methylene groups) while keeping the `well' (the 1,4$-$C$_{6}$H$_{4}$ moiety) constant.
 Accordingly, we have now synthesised new molecules and have determined the single
 molecule conductances of 1,4$-$HS(CH$_{2}$)$_{x}$C$_{6}$H$_{4}$(CH$_{2}$)$_{x}$SH
 (xPhx; x = 1, 3, 4, 6)
 using the $I(z)$ technique, and we have carried out transport calculations on the
 molecules. We find that the variation of molecular conductance with $n$ in this family
 is surprisingly small, and that this is the consequence of the presence of a
 Breit-Wigner resonance in the transmission curve (caused by the higher-energy of the
 two $\pi$ - bonding orbitals of the benzene ring). This resonance sharpens and moves
 closer to the contact Fermi energy as $n$ increases, offsetting to some degree the
 rapid exponential decrease in conductance with length that might be expected for
 increasing $n$. Thus, control over the conductance of these molecules can be exerted
 not only by substituents on the phenyl ring, but also by varying the alkyl chain length.
More importantly, we show that this system can be understood in terms of a simple
pair of back-to-back metal-semiconductor junctions where the alignment of
 the states in the internal part of the semiconductor primarily depends on how the
 electrostatic balance is established at the contact region. Such a balance is in turn
 mainly determined between the molecular mid-gap states and the metal Fermi level via
 charge transfer, which finally determines the position of the bulk resonance with
 respect to the Fermi level. We also show that investigating this system within this
 perspective is essential to understand the level alignment in the junction, which
 otherwise could not be fully explained by the initial trend in the gas phase.

\section{Experimental RESULTS}

\subsection{Syntheses, sample preparation and monolayer formation}
1Ph1 (as the free di-thiol) is commercially-available and was used as received.
 6Ph6 (di–thioacetate; Fig.~\ref{fig:fig1}) was synthesised as previously described
\cite{Leary2007} and the same general approach was used to synthesise 3Ph3 and 4Ph4 ;
 full details are provided in the Supplemental Material (SM) \cite{SM}.
Gold-coated glass substrates (Arrandee) of ca. 1 x 1 cm size were flame-annealed using
 a butane-air torch to red heat, to generate Au(111) terraces, \cite{Haiss1991} as
 confirmed by STM imaging (see SM \cite{SM}, Fig.1). The slides were allowed to cool, and were
 then immersed in a 10$^{-3}$ M solution of the appropriate di-thioacetate
 (3Ph3, 4Ph4, 6Ph6)
 or dithiol (1Ph1) in CH$_{2}$Cl$_{2}$ for 1-2 minutes (1Ph1; 3 minutes) to afford a
 partial monolayer coverage of adsorbed dithiol molecules. Polarisation modulated infrared
 reflection-absorption spectroscopy (PM-IRRAS) was used to confirm that the adsorption
 of the molecules took place (for an example, see Fig.2 in the SM), and STM imaging of
 these submonolayers showed bright spots indicative of the presence of molecule(s),
 particularly at step edges (see SM \cite{SM}, Fig.1c); molecular resolution could not be
 obtained, as is typical for highly-mobile adsorbed thiols at submonolayer coverages.

\subsection{Single molecule conductance determination}
The scanning tunnelling microscopy-based $I(z)$ technique was used to determine single
 molecule conductance values for 1Ph1, 3Ph3, 4Ph4 and 6Ph6. In this technique,
 tip-substrate contact is avoided.  A freshly-cut Au STM tip was brought into close
 proximity to the modified substrate surface using the appropriate set point current,
 and the tip was then retracted while the tunnelling current was measured. When a
 molecule (or molecules) bridge between tip and substrate, characteristic current
 plateaux in the current-vertical distance ($I-z$) traces result, and the molecular
 conductance is determined from a histogram of many such traces. In this work, at least
 500 $I-z$ traces showing current plateaux (defined as having length $\geq$ 0.5 nm) were
 combined into conductance histograms for any given experiment; typically, approximately
 15 $\%$ of $I-z$ traces showed a plateau. Of the remaining traces, approximately 60$\%$
 were straightforward exponential decays with no evidence of molecular bridge formation
 and the rest were noisy and showed no obvious plateau; $I-z$ traces in these categories
 were rejected. Two-dimensional histograms of conductance-vertical distance correlation
 were also constructed. The tip bias in all experiments was +0.6 V.

\begin{figure}[t]
\begin{center}
\includegraphics[width=8.2cm]{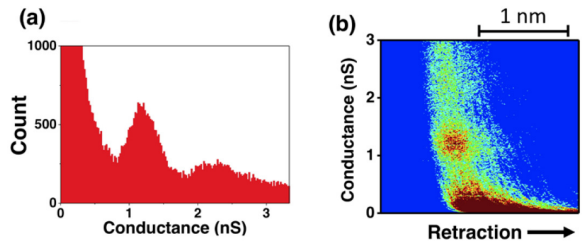}
\caption{(a) Histogram of 500 $I(s)$ scans obtained for 4Ph4 using a set point current
 of 10 nA, and (b) a 2-dimensional histogram plot, colour-coded to show frequency of
 occurrence of conductance as a function of distance in these $I(s)$ traces.}
\label{fig:fig2}
\end{center}
\end{figure}

\begin{figure}[t]
\begin{center}
\includegraphics[width=8.2cm]{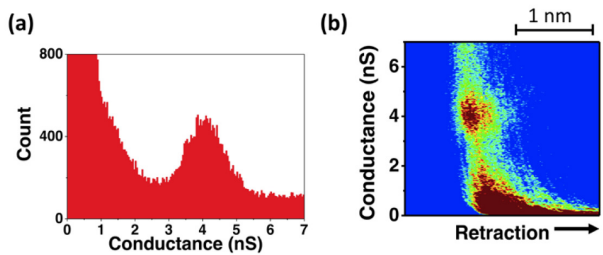}
\caption{(a) Histogram of 500 $I(s)$ scans obtained for 1Ph1 using a set point current
 of 30 nA, and (b) a 2-dimensional histogram plot, colour-coded to show frequency of
 occurrence of conductance as a function of distance in these $I(s)$ traces.}
\label{fig:fig3}
\end{center}
\end{figure}

Initially, we began by repeating the experiment of Leary et al.\cite{Leary2007} using
 6Ph6, with a set point current of 6 nA. This produced a relatively sharp histogram peak
 at a conductance of 0.82 $\pm$ 0.18 nS, in good agreement with the results reported by
 Leary et al (see SM \cite{SM}; Fig.4). A second clear peak can also be seen at approximately
 twice this value, corresponding to junctions formed with two molecules in the gap.
Next, we examined 4Ph4, 3Ph3 and 1Ph1. It was necessary to increase the set point
 current for these molecules, to 10, 20 and 30 nA respectively, since the molecules are
 progressively shorter, hence the tip has to be brought into closer proximity to the
 surface to ensure a reasonable percentage of successful junction formation events on
 tip retraction. Fig. ~\ref{fig:fig2} shows the results obtained for 4Ph4, and
 Fig.~\ref{fig:fig3}  those for 1Ph1; corresponding results for 3Ph3 are shown in
 Fig.3 in the SM \cite{SM}.
A small proportion of the $I-z$ traces from these experiments showed some evidence of a
 final, lower-conductance plateau, following the plateau that gave rise to the above
 conductance peaks. This suggested that lower-conductance junctions might form with
 these molecules at larger tip-substrate distances, but these plateaux do not give rise
 to a discernable separate conductance peak in either one-dimensional or two-dimensional
 histograms (see e.g. Fig. ~\ref{fig:fig2}(a) and ~\ref{fig:fig2}(b) where they are lost
 in the low-current noise at $<$ 0.8 nS). These events were quite rare, even when the
 set-point current was set to a smaller value to favour the formation of these longer,
 lower conductance plateaus over the shorter, higher conductance events, and so we did
 not analyse these events further.
The data for all experiments are compared in Table ~\ref{tbl:1}. 
We used a previously-published procedure to measure the break-off distance
 (the distance z at which the junctions break down)\cite{sedghi2008}  and to correct
 this for the initial set-point distance z$_{0}$ (the initial height of the STM tip).
 These break-off distances were similarly plotted as histograms (see SM \cite{SM}, Fig.6). Since, as is evident from Fig. 2 and 3, the conductances
 of the junctions are rather invariant with distance for these highly flexible molecules,
 we use the 95th percentile break-off distance as an estimate of the maximum length of
 the junction at breakdown, and it can be seen from Table 1 that this correlates well
 with the theoretical length of a junction between the two contacting gold atoms,
 approximated for the fully-extended molecular conformation using molecular mechanics
 (Spartan 14). The exception is 1Ph1, where the junctions at breakdown are somewhat
 longer than expected for a molecule of this length. It is possible that for a very
 short molecule, extension of the junction results in gold atoms being pulled out from
 the surface prior to junction breakdown.

\begin{table}
\begin{tabular}{ | c | c | c | c |}
\hline
Molecule & \begin{tabular}[c]{@{}c@{}} Conductance\\ (nS)\end{tabular} &
 \begin{tabular}[c]{@{}c@{}c@{}}95th percentile\\break-off   \\ distance (nm)\end{tabular}  &
 \begin{tabular}[c]{@{}c@{}c@{}c@{}}Calcd. Au...Au  \\
distance for \\ Au \textbar xPhx \textbar Au \\ junction (nm)$^{a}$ \end{tabular}  \\
\hline
1Ph1 & 4.14 $\pm$ 0.58 & 1.4 & 1.1 \\
\hline
3Ph3 & 2.01 $\pm$ 0.32 & 1.8 & 1.7 \\
\hline
4Ph4 & 1.21 $\pm$ 0.19 & 1.9 & 1.9 \\
\hline
6Ph6 & 0.82 $\pm$ 0.18 & 2.5 & 2.3 \\
\hline
\end{tabular}
\caption{Summary of conductance data for molecules xPhx. (a) Calculated for the
 fully-extended (all-$anti$ conformer) molecule, terminated by two Au-S bonds,
 using Spartan 14 molecular mechanics calculations.}
\label{tbl:1}
\end{table}

\begin{figure}[t]
\begin{center}
\includegraphics[width=8.2cm]{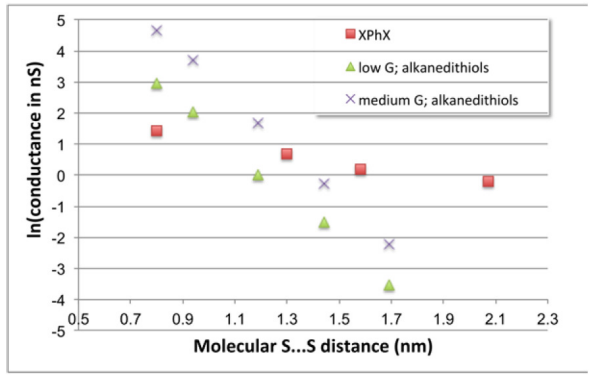}
\caption{Plots of ln(conductance in nS) vs. molecular length between the contact S atoms
 for the xPhx series (this work), and for the low and medium conductance groups as
 recently measured for 1,5-pentanedithiol, 1,6-hexanedithiol, 1,8-octanedithiol and
 1,10-decanedithiol in UHV\cite{Pires2013} and in ambient (1,12-dodecanedithiol\cite{yamada2008}). These examples are shown as these molecules have similar lengths to the shorter
 members of the xPhx series; the conductances of junctions with longer alkanedithiol
 have not been experimentally determined in this fashion as the conductance values would
 be too low.}
\label{fig:fig4}
\end{center}
\end{figure}

It is interesting to note that, alone of the molecules studied here, 1Ph1 has previously
 been studied by the STM-BJ technique, and at 0.6 V bias voltage, the conductance
 measured was 80 nS.\cite{xiao2004} It has previously been shown that the STM-BJ method
 favours the formation of high conductance molecular junctions, and this has been
 interpreted in terms of the contact atoms binding to gold at step edges or similar
 defects, producing a junction with a shorter metal | metal distance and with additional
 channels for conduction through the molecule due to the step edge gold atom in close
 proximity to the molecular backbone.\cite{haiss2009}
In Fig.~\ref{fig:fig4}, we compare the conductance as a function of length between the
 sulfur contact atoms (calculated for the fully-extended molecules using the molecular
 mechanics approach in Spartan14) for the xPhx series with recent data for the
 alkanedithiols, measured in UHV.\cite{Pires2013} A notable feature of the conductance data for the
 xPhx series is the very low apparent value of $\beta$ (calculated from the slopes of
 these lines; Eq.~(\ref{G-eq1})), 1.52 nm$^{-1}$. 

Since the only structural change in the molecules xPhx is the length of the alkyl chains,
 it is pertinent to compare the $\beta$ values with those found for alkanedithiols in
 related STM-based experiments; reported values range from 7-8.2 nm$^{-1}$ for all three
 conductance groups.\cite{haiss2009,li2007,busiakiewicz2010} Since the Fermi energy of
 the contacts has to fall between the HOMO and LUMO of the molecule in the junction, for
 molecules conducting via a tunnelling mechanism one expects the $\beta$ value to depend
 upon the HOMO-LUMO separation, with more conjugated molecules having a smaller HOMO-LUMO
 gap and hence a smaller $\beta$. For instance, for a series of oligophenyl methanethiols
 (C$_{6}$H$_{5}$CH$_{2}$SH, C$_{6}$H$_{5}$C$_{6}$H$_{4}$CH$_{2}$SH and
 C$_{6}$H$_{5}$C$_{6}$H$_{4}$C$_{6}$H$_{4}$CH$_{2}$SH), $\beta$ was determined as
 4.6 $\pm$ 0.7 nm$^{-1}$ using the GaIn/Ga$_{2}$O$_{3}$ eutectic contact technique on
 self-assembled monolayers of the molecules on template-stripped gold electrodes
 \cite{fracasso2013} (a similar value, 4.2 $\pm$ 0.7 nm$^{-1}$, was determined earlier
 using a conducting AFM technique\cite{wold2002}), while for a series of oligothiophenes
 with 5, 8, 11 and 14 2,5-coupled thiophene rings, $\beta$ was 1 nm$^{-1}$, measured by
 the STM break junction method.\cite{yamada2008} The lowest $\beta$ values so far
 measured have been for extended viologens (0.06 nm$^{-1}$) \cite{kolivoska2013} and for
 conjugated fused oligoporphyrins, oligoporphyrin-ethynylene and
 oligoporphyrin–butadiynylene systems, where values of 0.2 to 0.4 nm$^{-1}$ have been
 found, the exact value
 depending on structure.\cite{sedghi2012,sedghi2011,sedghi2008,li2012} It should be
 noted that although a low value of $\beta$ is often taken as an indication that the
 conductance mechanism could be hopping rather than superexchange (tunnelling), it has
 been shown that the conductance behaviour of even the most conjugated of these families
 of porphyrins as a function of temperature are still described well by theoretical
 models involving tunnelling.\cite{sedghi2012}
In this context, a value of $\beta$ of ca. 1.5 nm$^{-1}$ is remarkably small for the
 xPhx series, given that the only variable in this series is the number of methylene
 groups, for which we would expect $\beta$ to be 7-8 nm$^{-1}$. To check for a possible
 hopping mechanism, we have carried out conductance measurements as a function of
 temperature on a representative molecule from this series, 6Ph6 (for data, see Fig.5
 in the SM \cite{SM}). The conductance of this molecule as determined by the $I(z)$ technique did
 not vary, within experimental limits, over the temperature range 20-100 $^{\circ}$C.
 This confirms that a hopping mechanism is not operative, as expected since a single
 benzene ring is not redox-active within conventional potential limits.

\section{Theoretical modelling and discussion}
To gain insight into the conductance behavior of the xPhx system we performed DFT
 and electronic transport calculations with the help of the code Gaussian\cite{Gaussian}
 , using the PBE functional\cite{perdew1997} and the LANL2DZ basis set,\cite{wadt1985ab}
 unless otherwise specified. The molecular junctions were built by matching the S atoms
 of the molecule with the S atoms of the S-Au structure. The electronic transport
 calculations were carried out with ANT.G,\cite{alacant, Jacob2011} which is built as an
 interface to Gaussian. This code computes the electronic transmission using
 non-equilibrium Green's function techniques in the spirit of the Landauer formalism,
 employing parametrized tight-binding Bethe lattices in the far electrode description.
\cite{Jacob2011} A CRENBS basis set was used\cite{ross1990} for the Au atoms in the
 outermost layers. Further details of the calculations involving alternative contact
 sites (top, step) are given in the SM \cite{SM}.
We optimized the molecular structures for the geometries of 1Ph1, 3Ph3, 4Ph4 and 6Ph6
 in the gas phase and then relaxed a S atom on top of a 19 gold atom cluster (together
 with the gold atoms in the top layer). Then we analyzed their electronic structure
 after removing the terminal H atoms, which are known to be lost upon adsorption onto
 gold under our conditions.\cite{love2005} 
 
\begin{figure}[t]
\begin{center}
\includegraphics[width=8.2cm]{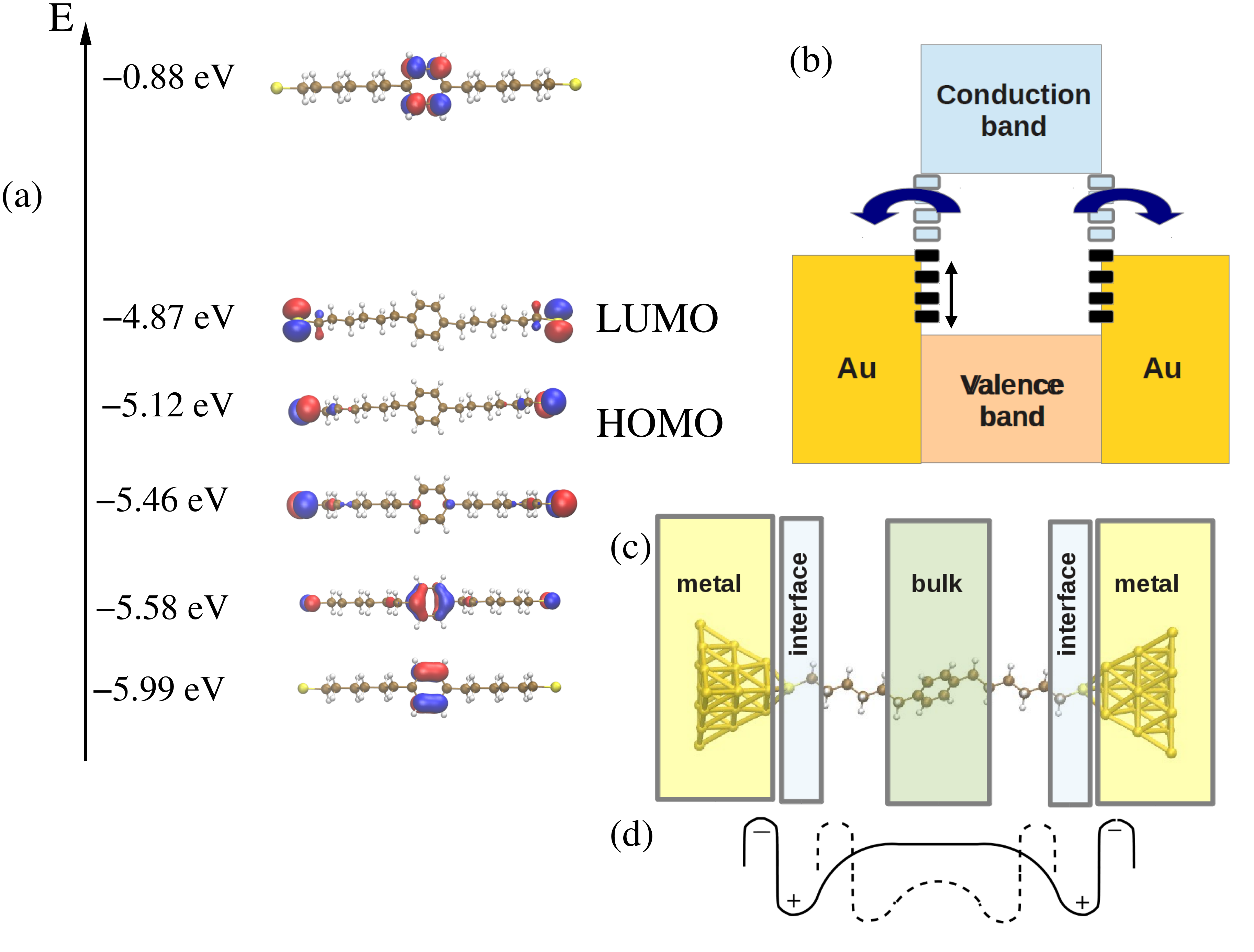}
\caption{(a) Frontier orbitals of 6Ph6; the H atoms of the thiols are omitted as these
 are lost on adsorption to gold;  (b) schematic picture of the metal-semiconductor-metal
 junction created once the molecule is connected to the electrodes. The vertical arrow
 indicates the energy difference between the metal Fermi level and the valence band;
 (c) A model of the 6Ph6 molecule in a
 junction, forming a back-to-back pair of metal-semiconductor interfaces where
 the benzene ring is
 treated as the semiconductor bulk, the gold contacts are the metal and the S and
 outermost C atoms represent an interface region. (d) Schematic electronic potential
 along the molecular junction depicted in panel (c) (solid line) and along a junction
 comprising a shorter molecule (dashed line). In the case of the longer molecule, the
 potential generated by the two point dipoles at the interfaces gradually vanishes in
 the bulk region, which is therefore only partially affected by the presence of the
 dipoles. This is not the case for the shorter molecule, which does not offer enough
 distance for this recovery and where the effect of the dipoles in the bulk region is
 stronger.
 }
\label{fig:fig5}
\end{center}
\end{figure}

 For all molecules, the HOMO-2 is localized
 on the benzene ring and is found to rise in energy as the chain length increases (see
 Fig. 9). As will be seen later in the discussion, the length dependence of
 the alignment of this orbital with the Fermi energy plays a determining role in the conductance behavior.
 The LUMO, HOMO and HOMO-1 all consist of spin degenerate levels, localized on the S
 atoms. Orbitals localized on the alkyl chains are far below in energy.  In the left
 panel of Fig.~\ref{fig:fig5}, we show the orbitals for the longest molecule (6Ph6)
 with the corresponding energies. It can be observed that the LUMO (localized on the
 S atoms) is quite close in energy to the HOMO, while the LUMO+1 (localized on the
 benzene) is far above.  More importantly, a clear spatial separation appears between
 the states localized on the S atoms and those localized on the benzene ring.

In our view, the LUMO+1 and HOMO-2 can be likened to the conduction-valence band structure in a
 semiconductor, while the states in between (LUMO, HOMO and HOMO-1) can be likened to
 the surface midgap states of the semiconductor as they are mainly localized on the S
 atoms that connect to the electrodes. Consequently, placing this molecule between
 two metal electrodes will form a junction that is expected to behave like a pair of
 back-to-back metal-semiconductor junctions  in which the chemical terminations
 of the semiconductor create states within its band gap (as schematically depicted 
 in Fig.~\ref{fig:fig5}b).
  For the purposes of this analogy we refer to the benzene core of the molecule as
 the ``semiconductor bulk''. The whole junction can then be thought of as divided into
 five main blocks: metal-interface-bulk-interface-metal (Fig.~\ref{fig:fig5}c).
 Each diode consists of a metal-(organic) semiconductor point-like junction. In our case, the
 semiconductor is intrinsic and the metal Fermi level falls between the HOMO and the
 LUMO of the central benzene ring (i.e. between the HOMO-2 and the LUMO+1 of the
 complete system). 
 
 It is not straightforward, a priori, to know where the Fermi level
 will lie in this gap. However, insights can be gained by considering what happens at
 the interface between a metal and a semiconductor.  In analogy with the semiconductor
 surface states, the states localized on the S atoms of the xPhx molecules should be
 strongly hybridized when interacting with the metal, giving rise to the so-called
 `gateway states'\cite{Widawsky2013}. The mismatch between the metal Fermi level and the
 molecular charge neutrality level (CNL) at the interface induces a charge transfer
 which in turn creates a point-like dipole;\cite{flores2009}  the energy levels in
 the inner part of the semiconductor bulk are then shifted up or down due to
 the dipole by a quantity which varies
 with the distance from the interface (Fig.~\ref{fig:fig5}d). The role of dipole
 interfaces has been extensively investigated in the case of organic monolayers or
 (more generally) single molecules lying flat on surfaces
 \cite{Heimel20064548,bokdam2011,hofmann2010,braun2009} but, to the best of our knowledge, a thorough study concerning their roles in junctions comprising a single molecular
 nanowire held between two metal clusters
 has not been explored to the same extent.\cite{gutierrez2003,Xue2001,peng2009}
 It is worth stressing that in this
 case, since the semiconductor is intrinsic, no depletion layer is formed. Consequently,
 variations in the energy position of the bulk levels must not be ascribed to the
 typical band bending associated with it in doped semiconductors. Here, the only
 electrostatic effect is related with the dipole established at the interface and will
 be confined to a narrow region close to it. Notice also that while in a planar junction the bands are
 shifted at (infinitely) large distances,\cite{leonard2000} in a single-point junction, such
 as the one considered here, the effect of the dipole effect is expected
 to decay rapidly along the transport direction. In what follows, we proceed to analyze
 the energetics of the four molecules when placed between two gold electrodes; we will
 show how studying the electrostatic balance established at the metal-molecule interface
 is essential for a thorough understanding of the level alignment, which in turn
 impacts on charge transport across the junction. 
 
 \begin{figure}[t]
\begin{center}
\includegraphics[width=8.2cm]{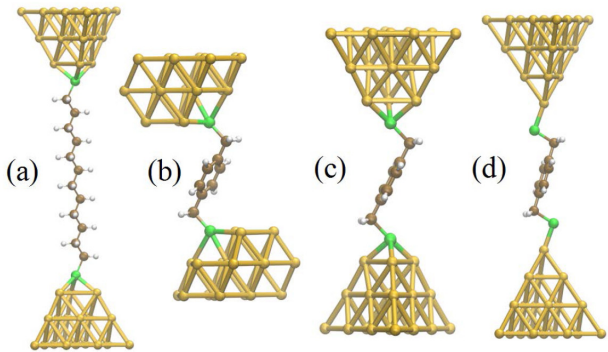}
\caption{Representative dithiols studied theoretically, showing different models for
 adsorption sites on gold. (a): dodecanedithiol bound at hollow sites, (b) 1Ph1 bound
 at step edge sites, (c) 1Ph1 bound at hollow sites and (d) 1Ph1 bound at top sites.}
\label{fig:fig6}
\end{center}
\end{figure}

\begin{figure}[t]
\begin{center}
\includegraphics[width=8.2cm]{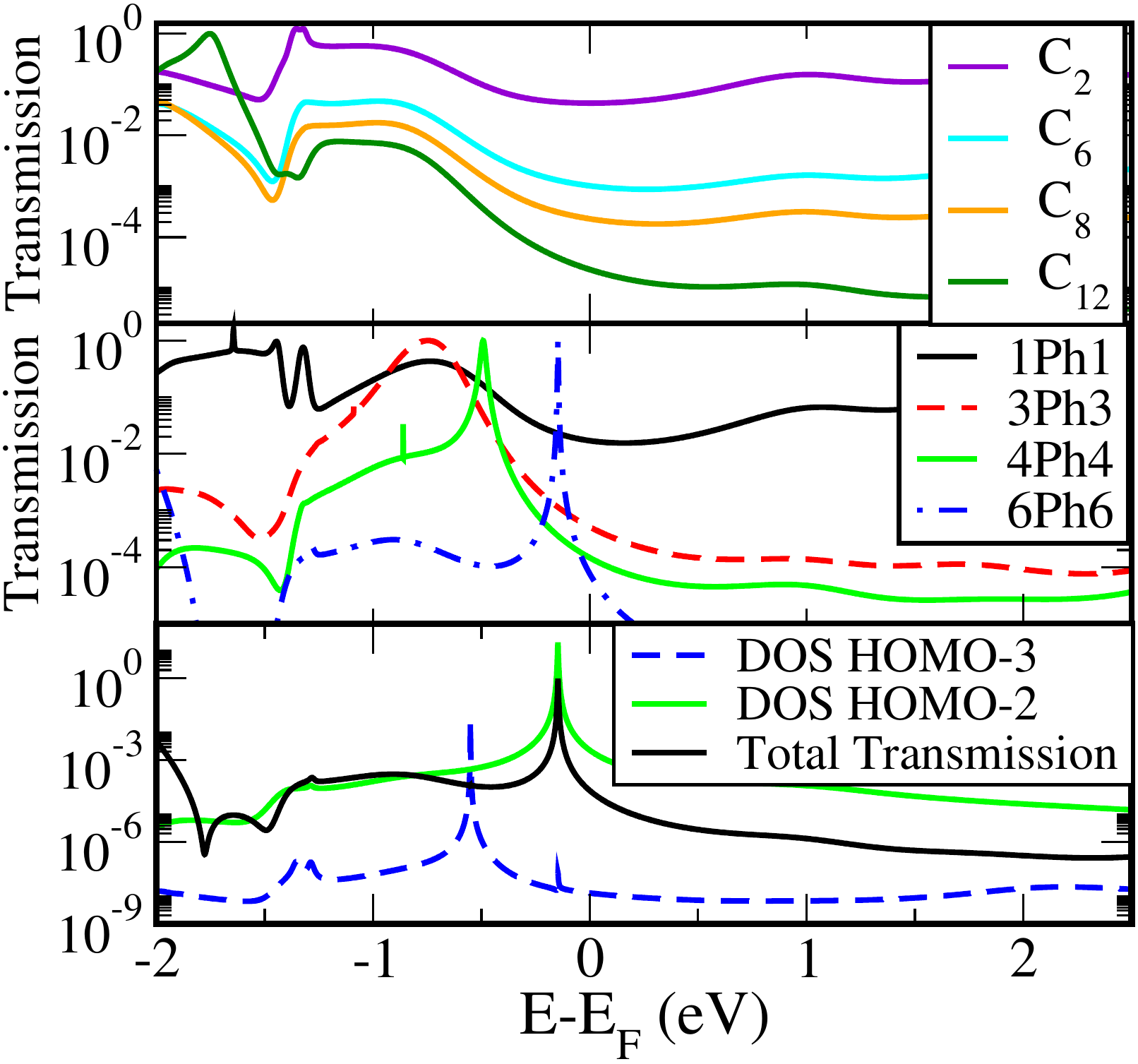}
\caption{Transmission as a function of energy for Cx  and xPhx in the hollow binding
 geometry and DOS projected onto HOMO -2 and HOMO-3 for 6Ph6 in the hollow geometry.
}
\label{fig:fig7}
\end{center}
\end{figure}

We constructed junctions in which each molecule is connected to two gold clusters, in
 the hollow, top or step binding geometry (see Fig.~\ref{fig:fig6}); however, in what
 follows, we will restrict discussion to the hollow position.
For comparison, we also studied a series of alkanedithiols of four different lengths
 (C2, C6, C8 and C12), also bound in the hollow position. Although their total lengths
 are different from the xPhx molecules, this comparison was made to give insight into
 the presence of any additional state(s) and into differences in the length dependence.
 In Fig.~\ref{fig:fig7}, we show the low-bias transmission curves obtained for the Cx
 and  xPhx
 series in the hollow geometry. For the xPhx molecules, the presence of additional
 peaks is evident. These Breit-Wigner resonances approach the Fermi level
 as the molecular length increases
 (consistent with previous calculations \cite{Huser,benesch2008}) and they do not appear
 in the case of the Cx molecules. The presence of these states creates an indentation
 in the barrier, confirming our earlier suggestion \cite{Leary2007}. A similar effect
 has previously been shown in another system.\cite{zotti2013} These resonances stem
 from the HOMO-2, which is localized on the benzene ring, thus they become narrower as
 the increasing benzene-gold distance decouples them from the gold contacts. To
some extent, the benzene ring behaves as a quantum dot, weakly coupled to the electrodes
by the alkyl chains. The
 trend in energy of the peaks approximately follows that in the gas phase for
 3Ph3, 4Ph4, and 6Ph6, whereas the corresponding peak of 1Ph1 lines up further down in
 energy (at -1.48 eV). Indeed, in the case of this shorter molecule, exceptional
 behavior is expected due to the strong hybridization of the orbitals.  In the case
 of the Cx molecules, the states closest to the Fermi level are localized mainly on
 the S atoms, while states localized on the whole chain appear only below -2 eV.

To discern the different contributions in the transmission curves of the xPhx molecules,
 we calculated the DOS projected onto the molecular orbitals. The results are illustrated
 for 6Ph6 in the lowest panel of Fig.~\ref{fig:fig7}. These projections clearly show
 that the sharp peak visible in the transmission curve close to the Fermi level for 6Ph6
 is due to the HOMO-2, while the HOMO-3, also mainly localised on the benzene ring,
 does not yield any peak in the transmission curve apart from a very small Fano resonance
 (interference effects in transport through benzene rings have been widely studied
\cite{darau2009}). The broad bumps at around -1 eV derive from the gateway states,
 i.e. from the hybridization of S-Au states at the interface. Notice that using the same
voltage 0.6 V as in the experiments is not expected to change the overall picture. To 
confirm this, we show, in the SM, a comparison between the transmission curves for 4Ph4
calculated at 0 and 0.6 V: the position of the HOMO-2 is not affected, while the bias
voltage seems to rather have an influence on the interface states.

We now turn to analyse these systems quantitatively. In what follows, we will focus on
 the metal-molecule interface of our system in order to discern the various contributions
 that determine the level alignment in the studied junctions. We
 again focus on the hollow binding geometry. To begin with, we have first analyzed the
 Mulliken charges in order to discern whether any significant charge transfer takes
 place in this system (see Table I in the  SM \cite{SM}). They show an increased electron charge
 (around 0.3) on each gold cluster and corresponding positive charge on the outermost
 carbon atoms, indicating charge transfer from the molecule onto the metal. 
Subsequently, we have analysed a system consisting of each molecule connected to a
 single Au19 cluster in a hollow position, with the aim of understanding the level
 alignment at just one metal-molecule interface. In this case, as well as the standard
 DFT calculation, we have also performed constrained DFT (c-DFT) calculations, as explained below. 
 By connecting a Bethe lattice to the metal cluster, a new system is built, where the
 molecule is now connected to a semi-infinite electrode (see Ref.~\onlinecite{Jacob2011}
 for more details).  The system can be hence divided into two parts, namely the lead
 and the main region that comprises the molecule and a certain number of metal layers.
 The density matrix of this region is calculated as

\begin{equation}
P(\mu)= -\frac{2}{\pi} Im \int^{0}_{-\infty}{dE G^{+}(E;\mu)}
\label{P-eq2}
\end{equation}
where the retarded Green's function is given by
\begin{equation}
G^{+}(E;\mu)=[(E-\mu)\cdot S-H-\Sigma(E,\mu)]^{-1}.
\label{Green-eq3}
\end{equation}
Here, $H$ is the Hamiltonian of the main region, $\Sigma$ is the lead self energy,
 which describes the coupling to the semi-infinite lead, and $\mu$ is a
 quantity by which the on-site energies of the Hamiltonian must be shifted in order to
 ensure the total-charge neutrality ($\mu$ is opposite in sign to the chemical potential). In our c-DFT any charge transfer between
 the two parts is forbidden. Both surface and molecule were forced to keep a number of electrons as to maintain charge neutrality. To this aim, $\mu$ was
 computed in alternation so as to meet the imposed charge constraints on the metal and the molecule.
 If $N_{S}$ and $N_{mol}$ are the desired number of electrons in the surface and the
 molecule, respectively, then two different chemical potentials $\mu_{S}$  and
$\mu_{mol}$  are calculated so that
\begin{equation}
Tr[P_S(\mu_{S})S_S]= \sum\limits_{i,j=1}^{NAO_{S}}P(i,j)S(i,j)=N_{S}
\label{Tr-eq4}
\end{equation}
\begin{equation}
Tr[P_{mol}(\mu_{mol})S_{mol}]= \sum\limits_{i,j=NAO_{S}+1}^{NAO_{S}+NAO_{mol}}P(i,j)S(i,j)=N_{mol},
\label{Tr-eq5}
\end{equation}
where
$NAO_{s}$ and $NAO_{mol}$ are the number of atomic orbitals of the surface and the
 molecule, respectively. 
%
%
%
At each iteration step the density matrix $P^{o}$ is then built as a block matrix as follows:
\begin{equation}
P^{o}= \begin{pmatrix} P_{S}(\mu_S) & P_{S-mol} \\ P_{S-mol} & P_{mol}(\mu_{mol}) \end{pmatrix}
\label{mat-eq8}
\end{equation}
The off-diagonal terms in the submatrices $P_{s-mol}$ were set to zero. The so-built
 density matrix is out of equilibrium, with two different chemical potentials in the
 metal and the molecule (at approximately -5.96 eV and -4.2 eV for metal and molecule,
 respectively), consequently giving rise to a step potential across the system.
The Fermi level of the molecule calculated in this way can be interpreted as the charge
 neutrality level (CNL) discussed in the literature, \cite{flores2009,Vazquez2005}
 but obtained in a more rigorous way.  It defines the direction in which electronic
 charge is most likely to be transferred. In our case, the CNL is definitely higher than
 the Au Fermi level, so charge transfer from the molecule onto the metal is expected,
 confirming what was indicated by the Mulliken charges in the junction. 

\begin{figure}[t]
\begin{center}
\includegraphics[width=8.2cm]{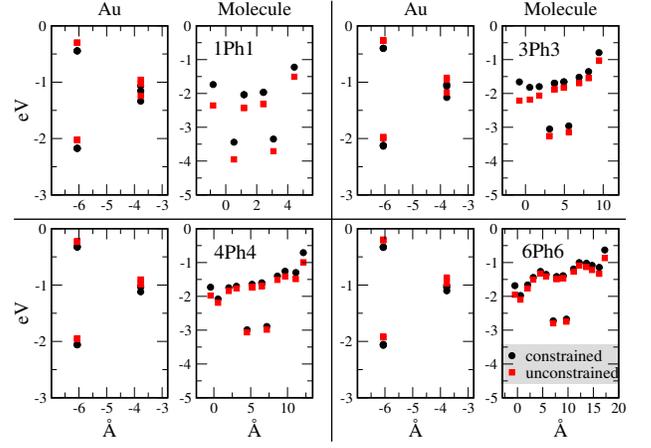}
\caption{On-site energies (averaged over the atomic orbitals of each atom) in the three
 innermost Au atoms and the molecular C atoms in the Au19-xPhx systems for all four
 molecules, in both constrained and unconstrained DFT.
}
\label{fig:fig8}
\end{center}
\end{figure}

We now turn to analyze what happens once the electronic charge has been transferred.
 To this end, we have computed the on-site energies along the Au19-molecule junctions
 and compared them with those resulting from the constrained calculation (the results
 are shown in Fig.~\ref{fig:fig8}). They were evaluated as the average (over the atomic
 orbitals of each atom) of the diagonal terms in the Kohn-Sham Hamiltonian. Since the
 absolute values along the vertical axes are not physically relevant, as they are
 strongly related to the basis-set choice for each atom, they have been offset for
 clarity. Hence, we will only focus on the comparison between the values obtained in
 two kinds of calculations.   At the interface, we see a clear sign of the expected
 dipole; in the unconstrained calculation the values are shifted upwards in the metal
 and downwards in the molecule. In the latter, the effect is stronger in the carbon
 atoms closer to the metal, fading away as the distance from the metal increases.
 In 1Ph1, the
 shortest molecule, the benzene ring is quite close to the gold and shows some
 positive charge (0.1) in the unconstrained calculation, as compared to that constrained.
 Indeed, for 1Ph1 the on-site energies are strongly affected also in the C atoms of the
 benzene ring. It is also worth adding that, when a second cluster is added at the other
 end of the molecules to complete the junctions, the effects caused by the dipole in
 the bulk region are expected to increase. The Mulliken charges along the Au19-molecule
junctions in the constrained and unconstrained calculations are shown in the SM.\cite{SM} 

\begin{figure}[t]
\begin{center}
\includegraphics[width=8.2cm]{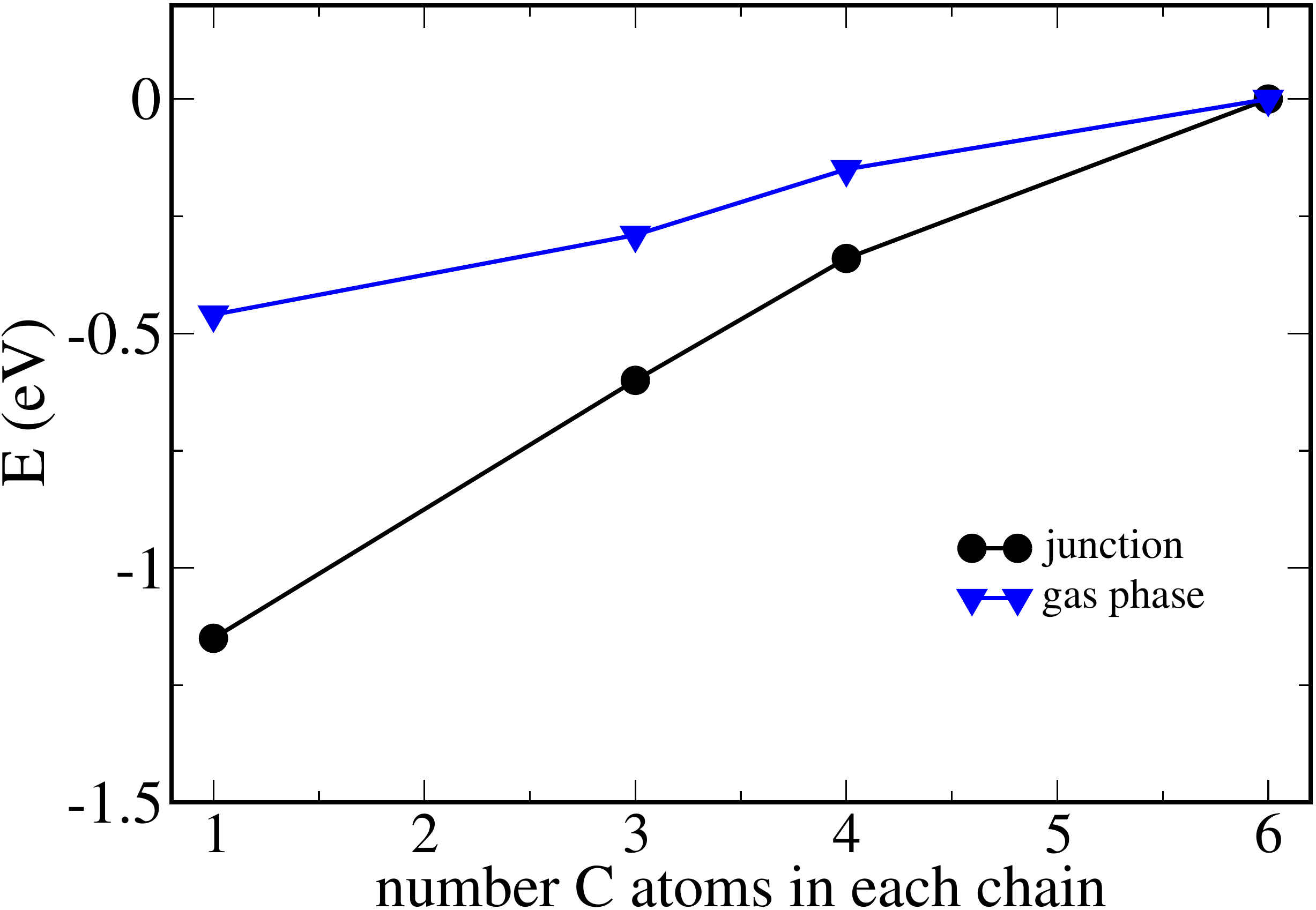}
\caption{Energetic position of the HOMO-2 for each molecule in the junction and in the
 gas phase.  All values have been shifted to set those for 6Ph6 to zero.  It can be
 seen that the original displacement of the HOMO-2 in the gas phase increases in the
 junction.
}
\label{fig:fig9}
\end{center}
\end{figure}

We summarize all these results as follows: when the xPhx molecules approach the metal,
 Fermi-level alignment is achieved upon charge transfer from the sulfur, or the
 outermost carbon atoms, onto the metal. This generates a dipole potential mainly near the
 interface. The molecular bulk states (HOMO-2 levels) then behave like spectators,
 with their energetic positions being dragged down under the influence of the dipole
 potential according to the distance from the interface (Fig.~\ref{fig:fig5}d), and
 never crossing the Fermi level (see, for example, the case of the top binding geometry
 for the hypothetical 8Ph8 in the SM \cite{SM}). The level alignment of the bulk states is
 governed purely by electrostatics because of the clear separation from the interface
 (apart from 1Ph1) as already discussed. Note that this would not be possible in fully
 conjugated molecules, where other effects such as hybridization would also play a role. Notice that any electrostatic-induced shift is added to the one the HOMO-2 already
 presents in the gas phase for increasing chain length (see, in Fig.~\ref{fig:fig9},
 a comparison between the energy shift experienced by this orbital in the gas phase
 and in the junction).

It should be recalled that there is some uncertainty in the position of the HOMO-2 peak
 in the transmission curves as it corresponds to a Kohn-Sham state, and it might be
 located at lower energies. Nevertheless, the computed transmission curves provide an
 insight into the alignment dynamics, i.e. they clearly show how the alignment of the
 bulk states is governed by the gas-phase behavior, modulated by what happens at the
 interface. 
To support our interpretation of the numerical results, it is important to verify that
 the benzene ring falls inside the space region in which the dipole-potential effect is
 confined. To this aim, we have calculated the potential generated in the middle of the
 benzene ring by the charges at the interface. The results are reported in Table II of
 SM \cite{SM} and show decreasing values for increasing chain length, with maximum values of
 around 1 eV.  

\begin{figure}[t]
\begin{center}
\includegraphics[width=8.2cm]{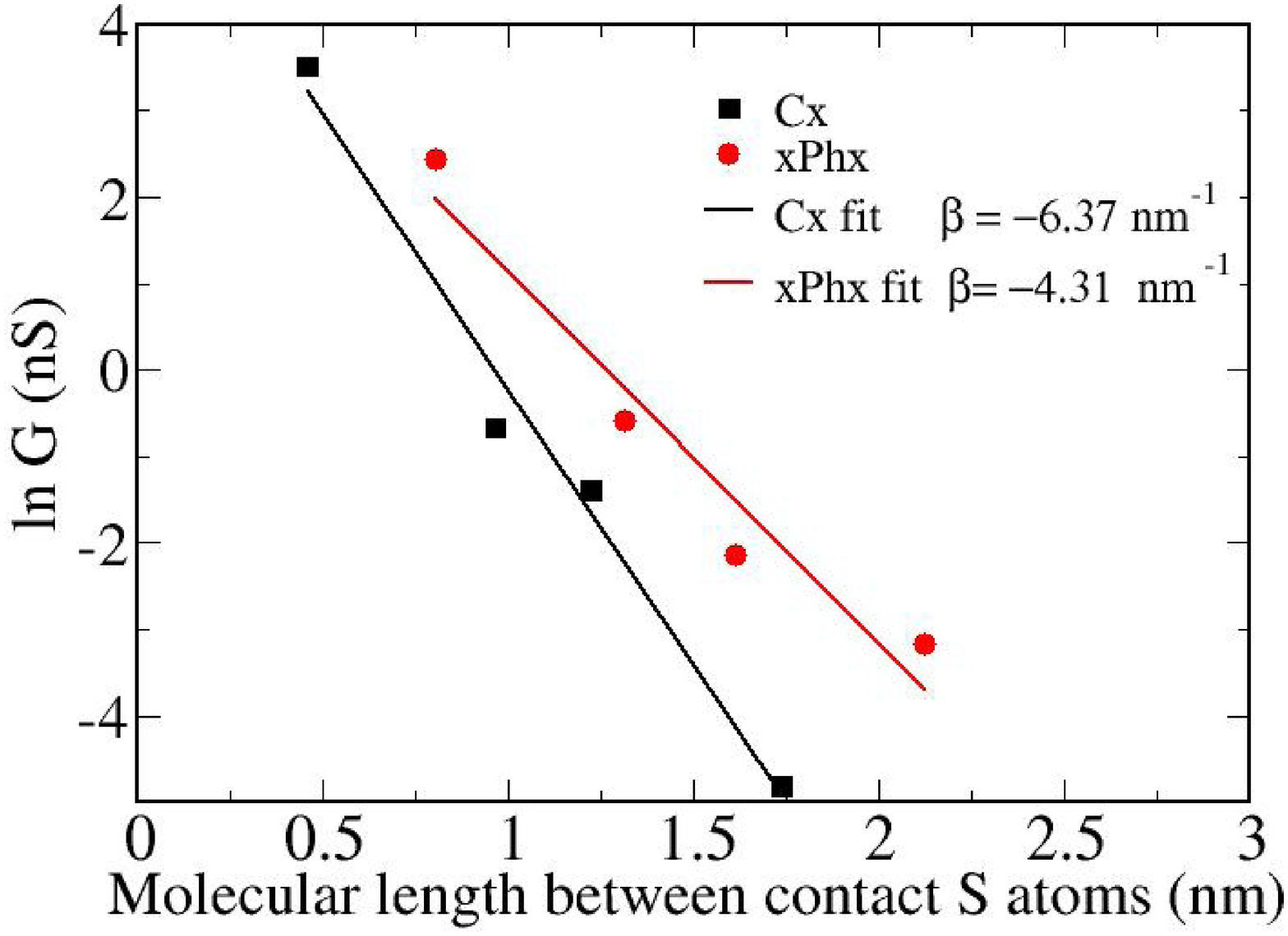}
\caption{The calculated ln(conductance) as a function of the molecular length for Cx and xPhx in the hollow binding geometry}
\label{fig:fig10}
\end{center}
\end{figure}

Overall, the scenario described above provides a good rationalisation for the
 experimentally observed low value of $\beta$. There are two counteracting effects,
 namely the expected decay due to increasing molecular length on the one hand, together
 with the increasing sharpness of the HOMO-2 resonance in the molecules with longer
 polymethylene bridges, and the approach to the Fermi energy of the resonance peak on
 the other hand. Fig.~\ref{fig:fig10} shows a plot of calculated zero bias conductance
 vs. molecular length for the xPhx and Cx molecules. Although the absolute values of the
 conductances differ substantially from the experimentally observed values, it can be
 clearly seen that the calculated $\beta$ value is smaller for the xPhx series than for
 the Cx molecules, although the factor by which they differ (ca. 1.5) is substantially
 smaller than that seen experimentally (ca. 4.8). Nevertheless, given the known problems
 with DFT, and the limitations inherent in the modeling of the contacts, we believe
 that this offers a good explanation for the experimental observations.
Ultimately, possible asymmetries in the experimental junctions should be taken into
 account (not considered here) which lead to transmission resonances with a maximum
height lower than 1. While this is
 not expected to change the conductance values for the Cx molecules, for which the
 resonances show up further below the Fermi energy, it might change the conductance
 values for the xPhx molecules and, consequently, the $\beta$ factor.

\section{CONCLUSIONS}
We studied the series of molecules
 HS(CH$_{2}$)$_{x}$[1,4–C$_{6}$H$_{4}$](CH$_{2}$)$_{x}$SH (x = 1, 3, 4, 6) and
 found that the rate of  conductance decay with molecular length is smaller than that
 measured previously for alkanedithiols. This was found to be due to the decreasing
 distance in energy of the Breit-Wigner resonance from the Fermi level for increasing
 molecular length. This trend originates from the orbital behaviour in the gas phase but
 is modulated by the electrostatic balance established at the electrode contact region.

\section{Acknowledgments}
This research was supported by the EPSRC (Grant No.
EP/H035184/1), by MINECO under Grant No. FIS2013-47328,
 by the European Union structural funds and the
Comunidad de Madrid MAD2D-CM Program under Grant.
P2013/MIT-2850, and by Generalitat Valenciana under Grant
PROMETEO/2012/011. We thank M. Soriano for help with the
computational part and acknowledge the computer resources,
technical expertise, and assistance provided by the Centro
de Computacion Cientifica of the Universidad Autonoma de
Madrid.


\bibliography{literature}
\end{document}